\documentclass{article}

\usepackage{multirow}
\usepackage[table,xcdraw]{xcolor}
\usepackage[utf8]{inputenc}
\usepackage{authblk}
\usepackage{abstract}
\usepackage{url}
\usepackage{geometry}
\usepackage{graphicx}
\usepackage{amsmath,amssymb}

\usepackage{url}

\title{DAOs of Collective Intelligence? Unraveling the Complexity of Blockchain Governance in Decentralized Autonomous Organizations}

\author[1,2]{Mark C. Ballandies}
\author[3]{Dino Carpentras}
\author[4]{Evangelos Pournaras}

\affil[1]{University of Zurich, Zurich, Switzerland}
\affil[2]{WiHi, Zug, Switzerland}
\affil[3]{ETH Zurich, Zurich, Switzerland}
\affil[4]{School of Computing, University of Leeds, Leeds, UK}

\date{February 26, 2025}

\begin{document}
	\maketitle
	
	\begin{abstract}
		Decentralized autonomous organizations (DAOs) have transformed organizational structures by shifting from traditional hierarchical control to decentralized approaches, leveraging blockchain and cryptoeconomics. Despite managing significant funds and building global networks, DAOs face challenges like declining participation, increasing centralization, and inabilities to adapt to changing environments, which stifle innovation. This paper explores DAOs as complex systems and applies complexity science to explain their inefficiencies. In particular, we discuss DAO challenges, their complex nature, and introduce the self-organization mechanisms of collective intelligence, digital democracy, and adaptation. By applying these mechanisms to refine DAO design and construction, a conceptual framework for assessing a DAO’s viability is created. This contribution lays the foundation for future research at the intersection of complexity science, digital democracy and DAOs.
	\end{abstract}
	
	
	\vspace{1em}
	\noindent\textbf{Corresponding author:} \texttt{markchristopher.ballandies@uzh.ch}

\section{Introduction}

Managing and regulating complex systems through hierarchical, top-down control approaches is challenging, often resulting in reduced system sustainability and resilience \cite{helbing2021next}. In contrast, bottom-up and decentralized methods such as collective intelligence, digital democracy, and adaptation have proven more effective in handling such complexity \cite{helbing2023democracy,Pournaras2020, carpentras2024empowering}. These approaches are successfully utilized by policymakers, such as in the city of Aarau \cite{Welling2023,yang2023designing}, and are also being implemented in the corporate world. This ranges from self-organized teams or projects within hierarchical structures \cite{hoda2016multi,hoda2010organizing,spychiger2023organizing} to completely decentralized organizations \cite{laloux2014reinventing,wyrzykowska2019teal,Majumdar2023,hunhevicz2024decentralized}.

Decentralized \textit{autonomous} organizations (DAOs) are blockchain-based organizations that accelerate this observed decentralization trend in society. Governed democratically through participatory algorithms, DAOs transcend geographical, cultural, and traditional boundaries, enabling global membership and cooperation to the thousands \cite{wright2021rise,Pournaras2020}. In particular, they extend the concept of self-organizing organizations by removing the clear boundary between the inside and the outside of the organization, creating permeable and flexible borders via which anyone can quickly start influencing the decision-making of the organization and participate in the execution of tasks \cite{spychiger2023organizing}. 

DAOs have proven their efficacy by managing significant assets, upwards of \$500 million, and showcasing capabilities for rapid capital deployment \cite{wright2021rise}. 
Beyond capital pooling, DAOs also excel in operational efficiencies. Decision-making processes are enhanced through algorithmic systems or blockchain technology, leading, for instance, to more efficient and cost-effective voting mechanisms~\cite{wright2021rise,Pournaras2020}. Regular, ongoing voting replaces traditional periodic voting, making it economically viable for members to play a more active role in organizational management~\cite{wright2021rise}. In general, smart contracts facilitate liquid democracy by lowering the costs and complexities of proxy-based voting systems and enhancing the transparency of decision-making. These features aid in internal control, reducing fraud and the need for constant monitoring \cite{wright2021rise}.
These characteristics position DAOs as a novel organizational model, offering a mix of enhanced democratic participation, operational efficiency, and investment capabilities, fundamentally reshaping how organizations are managed and operate.

Despite their potential, DAOs face significant challenges, including declining participation \cite{wright2021rise,rikken2019governance,faqir2021comparative}, increasing centralization \cite{santana2022blockchain}, adaptation difficulties~\cite{santana2022blockchain} and balancing contradicting instantiated values \cite{hunhevicz2024decentralized}. 
This paper examines DAOs as complex systems and applies principles of complexity science and digital democracy to address these challenges which are outlined in greater details in Section \ref{sec:dao}. By examining self-organization mechanisms such as collective intelligence, digital democracy, and adaptation (Section \ref{sec:complex}), Section \ref{sec:dao_framework} introduces a conceptual framework for assessing DAO viability, providing a foundation for analyzing DAOs and advancing research at the intersection of complexity science, digital democracy, and DAOs. In particular, the conceptual framework \cite{jabareen2009building} interlinks the three DAO dimensions—Community, Digital Democracy, and Adaptation—which together comprehensively inform a DAO’s viability. The paper concludes with a summary and outlook in Section \ref{sec:conclusion}.

\section{Background}

\subsection{Decentralized autonomous organizations (DAOs)}
\label{sec:dao}
A DAO consists of a community composed of individuals with shared relationships and varying goals. The DAO arises when these individuals collaborate towards a shared vision \cite{ospina2023daocommunity}. When properly balanced, the synergy between community and organization enables extraordinary value flows~\cite{ballandies2023onocoy}, such as, for instance, establishing global infrastructure networks~\cite{jagtap2021federated,ballandies2023taxonomy}.
DAOs are blockchain-based organizations, ranging from those using smart contracts for governance and ownership \cite{bellavitis2023rise,wright2021rise} to networks like Bitcoin, which operate without them \cite{hsieh2018bitcoin,ballandies2025bitcoin}. The following sections illustrate their core mechanisms and challenges.

\subsubsection{Mechanisms:}

DAOs utilize several mechanisms to manage and steer their activities.
The most utilized mechanism is the Improvement Proposal that was pioneered by Bitcoin. Community members draft (technical) documents which once accepted become binding to all community members. For instance, the three largest DAO platform utilize such proposal driven mechanisms \cite{faqir2021comparative}.
In the case of Bitcoin, no traditional voting is applied on these improvement proposals, but community members autonomously decide if they follow them, eventually being accepted when the largest hash power supports a proposal \cite{fish2024AnalyzingBitcoin,ballandies2025bitcoin}. This resembles voting by the feet as observed in democracies \cite{banzhaf2008people}.  

However, most DAOs apply explicit decision-makings systems to agree on those proposals \cite{faqir2021comparative} such as varying voting mechanisms \cite{wright2021rise}.
These approaches have off-chain (digital hubs, social media, and other instruments) and on-chain (includes consensus mechanisms, smart contracts and intelligent matching) governance processes \cite{santana2022blockchain}.
It has been found that the type of mechanism has an impact on the percentage of accepted proposals within a DAO \cite{faqir2021comparative}.
Often blockchain-based tokens, referred to as governance tokens, are utilized to implement those decision-making mechanisms \cite{wright2021rise,Li2024,ding2023survey}.

DAOs utilize participatory input from a large and wide group of stakeholders~\cite{wright2021rise} referred to as the community. The borders to enter the DAO community are permeable and membership can be short-term. Also, members of a DAO usually have equal access to information concerning the DAO~\cite{wright2021rise}. Though members are often enabled to participate anonymously \cite{spychiger2023organizing}, a core element of DAOs is that the mechanisms governing the DAO and the important information within (e.g. vote results) are made transparent \cite{wright2021rise}. 
In the event of irreconcilable disagreements, DAOs employ forking: a faction splits off using novel mechanisms incompatible with the rest of the community~\cite{wright2021rise}. Forking is possible because community members have the autonomy to determine the defining attributes of their DAO and the authority to utilize its tools and mechanisms as they see fit, even if other fractions of the community hold an opposing views.

\subsubsection{Challenges}
\label{sec:dao_challenges}
DAOs, though attempting to be decentralized, are often exhibiting influences from centralized stakeholders. For instance, because decision-making power is often tied to token ownership, and founders, investors, and exchanges usually control a larger share of tokens compared to average community members, DAOs often create plutocracies rather than true democracies \cite{pena2023categorization,feichtinger2023hidden,de2022structural}. Also, besides token concentration, developers who launch DAOs can exercise implicit control due to their deeper understanding of the underlying principles and smart contract rules of a DAO \cite{santana2022blockchain}. 
DAOs also exhibit high approval rates for proposals, possibly implying that DAO members tend to only present proposals they believe pass which could suggest a lack of experimentation with new ideas \cite{faqir2021comparative}.
In this context, while smart contracts aim to streamline decision making, the complexity and time required for consensus-building can hinder efficient action-taking and dissuade participation in DAOs \cite{wright2021rise,rikken2019governance}. This raises questions about the comparative efficiency of DAOs against more hierarchical organizations \cite{wright2021rise}. Participation in DAOs exhibits a declining trend in average voter turnout \cite{faqir2021comparative}, suggesting that DAOs may not effectively enable meaningful member contributions. However, research on digital democracy and referendum turnouts in Switzerland suggests that cumulative voter participation is a more appropriate measure, as individuals may not vote in every referendum but still engage in some \cite{serdult2013partizipation,serdult2021referendum}.



In summary, DAOs face challenges 
when it comes to centralization and the ability to act, which are significant obstacles to their long-term stability \cite{santana2022blockchain}. 
In the rest of this article, we illustrate how these challenges can be approached by considering findings from complexity science.

\subsection{Complex Systems and self-organization}
\label{sec:complex}


Complex systems are systems in which connections between components play a key role \cite{ladyman2013complex}.
 For such reason, it is often said that complex systems are more than the sum of their parts \cite{beckage2013more} because they exhibit properties that cannot be understood by examining individual components in isolation. For example, the emergent flocking behavior of birds only manifests when they interact, not when studied individually. In particular, these systems display emergent properties at the macro-level (e.g. flock) which differ significantly from those at the micro-level (e.g. bird). Additionally, complex systems exhibit chaotic behavior \cite{heylighen2001science,helbing2021self}, such as the butterfly effect, where minor changes can cause major impacts, or conversely, significant changes may have minimal overall effects.
The amount of interconnectivity of the current world makes it extremely complex. On top of this, new data and new connections are constantly generated resulting in a combinatorial growth in complexity. 
Complicating matters further, the emergent system properties have feedback effects on the individual parts, further increasing complexity \cite{helbing2021self}.
Therefore, hierarchical top-down control and decision-making fail in complex systems, as a single node cannot predict future behaviors or control the attainment of specific future states \cite{heylighen2001science}. Consequently, effective management of complex systems requires bottom-up, decentralized approaches rather than top-down control \cite{heylighen2001science,helbing2021next,helbing2023democracy,Pournaras2020}.

\begin{figure}[tbh]
\centering
\includegraphics[width=5in]{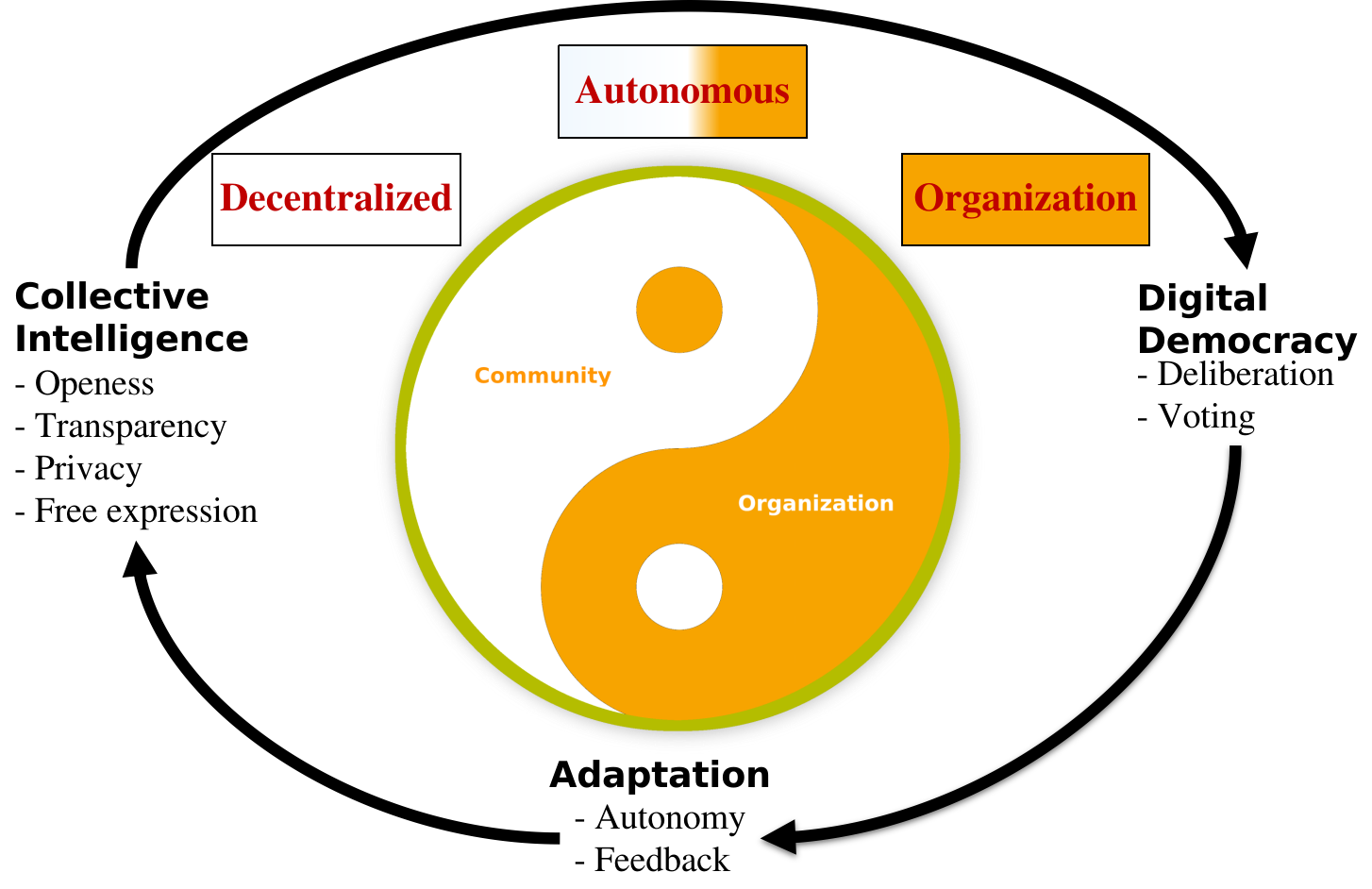}
\caption{DAO Viability Framework: Collective Intelligence, Digital Democracy, and Adaptation and their principles, each tied to a DAO's Decentralization, Autonomy, and Organization. Balancing these mechanisms is essential for a DAO to thrive.}
\label{fig:daoframework}
\end{figure}

The following three bottom-up, self-organization mechanisms have been identified to effectively control and manage complex systems:
\subsubsection{Collective intelligence} Collective Intelligence (CI) \cite{woolley2010evidence}, or wisdom of the crowds \cite{surowiecki2005wisdom}, encapsulates the enhanced ability of groups to engage in decision-making and brainstorm, with greater efficacy than individuals acting alone \cite{galesic2023beyond,suran2020frameworks,centola2022network}. The interest in CI has seen a notable rise in recent times, a development propelled by empirical research \cite{wolf2015collective} as well as progress in areas such as agent-based computational models \cite{reia2019agent,singh2009agent,singh2009agent2}, analysis of social networks \cite{centola2022network,ha2022collective}, and artificial intelligence \cite{weld2015artificial,singh2009artificial,jung2017computational}. This body of work suggests that under certain conditions, collective inputs from non-experts can surpass the performance of experts~\cite{hong2004groups}.
Nevertheless, there are also cases where a group of people can perform worse (and thus be less smart) than its individual components (sometimes called "collective madness") \cite{waddington2008madness}. 
In general, CI has several preconditions that must be fulfilled for it to be established:
a) \textit{Diversity} among members is essential for the emergence of collective intelligence in groups and societies \cite{woolley2010evidence,helbing2023democracy,mann2017optimal}. The larger and more diverse the group, the greater the potential for improved collective intelligence. b) \textit{Transparency} is required for group members to search for suitable information for a given task \cite{helbing2022socio,helbing2023democracy,helbing2021networked}. c) \textit{Privacy} in information exploration and solution development is necessary to prevent individuals from being externally manipulated to favor a particular solution \cite{helbing2022socio}. d) \textit{Freedom of expression} for sharing relevant information and solution approaches is key to faciliate learning~\cite{helbing2022socio,helbing2023democracy}


\subsubsection{Digital Democracy} Digital Democracy extends collective intelligence to decision-making in organizations by combining it with deliberation and voting mechanisms \cite{Helbing2015,de2003democracy}.
Based on the information from the collective intelligence, digital democracy adds two further steps \cite{hanggli2021human,helbing2022socio,helbing2023democracy,yang2025bridging}: a) \textit{Deliberation} involves the integration of various perspectives and solutions through a deliberative process to create a set of aggregated solutions that serve the group effectively. b) \textit{Voting} is then applied by the group on the aggregated solutions. Voting consists of two parts: i) input method defines the way how voting is conducted (e.g. preferential voting \cite{hare1873election,posner2017quadratic,Welling2023}), and ii) the aggregation method defines how the votes are counted. 


\subsubsection{Adaptation} Adaptation refers to the ability of complex systems to efficiently adjust to their environment without centralized coordination or control. Examples from nature include the human nervous system, bird flocks, and fluid crystallization, which operate without central authority, relying solely on local interactions and feedback to form robust structures \cite{heylighen2001science,kephart2003vision}.
In techno-socio contexts, adaptation can similarly enhance efficiency. For instance, leveraging local interactions and feedback mechanisms has been shown to significantly reduce average waiting times at traffic lights compared to centralized, top-down control~\cite{lammer2008self}.

Efficient adaptation in complex systems requires granting components autonomy to adjust based on local knowledge and employing feedback mechanisms to refine actions through impact assessment~\cite{heylighen2001science,brueckner2006organization}. Specifically, adaptation depends on:
a) \textit{autonomy to act}, enabling independent responses; and
b) \textit{real-time feedback}, which may accelerate development (positive feedback) or promote stability (negative feedback).

\subsection{A DAO as a Complex System}

A DAO can be modeled as a set of $N$ agents, each indexed by $i \in \{1, \ldots, N\}$, and each holding a state $s_i(t) \in \mathbb{R}$ at time $t$. The community relationships within the DAO can be modelled by an adjacency matrix $A = [a_{ij}]$, where $a_{ij} \neq 0$ indicates an interaction between agent $i$ and agent $j$. At each discrete time step the agents state is updated,
\[
s_i(t+1) \;=\; f\!\Bigl(s_i(t),\, \{s_j(t)\mid a_{ij}\neq0\}\Bigr),
\]
where $f(\cdot)$ is a local update rule reflecting DAO-specific (organizational) mechanisms (e.g.\ voting, resource distribution). A global performance measure $C(t)$ can then be defined to capture a DAOs overall state:
\[
C(t) \;=\; \frac{1}{N}\sum_{i=1}^{N} g\bigl(s_i(t)\bigr),
\]
with $g(\cdot)$ measuring each agent's contribution (e.g. to a vote).

Helbing \cite{helbing2012social} demonstrated how such networks of agents governed by local interaction rules can lead to self-organization. In particular, DAOs in this formulation exhibit nonlinearities, phase transitions, and emergent collective behavior that cannot be explained solely by individual behaviors \cite{helbing2012social}, thus making them complex systems.

For instance, let each agent $i$ holds a binary state $s_i(t)\in\{A,B\}$ at time $t$, representing its willingness to adopt competing standards, such as a block size in a blockchain network. 

\noindent
  Define the willingness of an agents neighborhood to adopt the standards as:
\[
x^A_i(t)=\frac{\sum_{j=1}^N a_{ij}\,\mathbf{1}[\,s_j(t)=A\,]}{\sum_{j=1}^N a_{ij}}, 
\quad
x^B_i(t)=\frac{\sum_{j=1}^N a_{ij}\,\mathbf{1}[\,s_j(t)=B\,]}{\sum_{j=1}^N a_{ij}}.
\]
Given parameters $q\in[0,1]$ (relative advantage of one option over the other) and switching costs $c_A, c_B\ge0$, the local update rule for an agent $i$ is:
\[
s_i(t+1) =
\begin{cases}
	B, & \text{if } s_i(t)=A \text{ and } x^A_i(t)<1 - q - c_A,\\
	A, & \text{if } s_i(t)=B \text{ and } x^B_i(t)<q - c_B,\\
	s_i(t), & \text{otherwise}.
\end{cases}
\]

This model has shown that competition between several (technological) options gives rise to bi-directional (or, more generally, multi-directional) percolation processes \cite{roca2011percolate}. Mapped onto a DAO, multi-percolation thus provides a powerful theoretical tool to understand, for example in the case of block size adoption, how forking may emerge as the coexistence of two technological standards within a community that eventually splits. It also informs strategies aimed at securing the adoption of a singular solution (e.g., ensuring only one standard percolates).

\section{DAO viability framework}
\label{sec:dao_framework}
This work defines a Decentralized Autonomous Organization (DAO) by its three key concepts. First, it is \textit{decentralized}, characterized by a broad community of interacting peers. Second, it is \textit{autonomous}, with community members operating independently of external or internal coercive forces. Third, it is \textit{organized}, possessing mechanisms for coordination and decision-making.

Considering DAOs as complex systems that require decentralized control (Section \ref{sec:complex}, we use this definition to position the three self-organization mechanisms (Section~\ref{sec:complex}) within the domain of DAOs, effectively guiding the design of well-functioning and viable DAOs. In particular, a DAO must instantiate the principles of the three self-organizing mechanisms for effective functioning, as illustrated in the proposed DAO Viability Framework (Figure \ref{fig:daoframework}).
Thus, taking a complexity science perspective, a DAO is a complex system that when it functions well harnesses the collective intelligence of its decentralized community, applies it through digital democracy for decision-making, and then capitalizes on the autonomy of its members to implement these decisions in an adaptive way in responses to environmental changes.

Table~\ref{tab:dao_framework} provides a detailed overview of the principles introduced in Section \ref{sec:complex} that must be met for the three self-organization mechanisms to function effectively.
\begin{table}[]
\caption{DAO viability framework: The eight principles of the three self-organization mechanisms are required to be instantiated within a DAO for it to be effective and viable. The framework supports the evaluation of DAOs by analysing the accomplishment of the individual principles, as presented for the MetaDAO (last column).}
\begin{tabular}{llll} \hline
\textbf{}                                                                            & \textbf{Princip.}                                        & \textbf{Goal}                                                                                                                                                            & \textbf{\begin{tabular}[c]{@{}l@{}}Accomp \\ lishment\end{tabular}} \\  \hline          
                                                                                     & Diversity                                                   & 
                                                                                     \begin{tabular}[c]{@{}l@{}}Permeable borders enable diverse actors to enter and \\ participate in the community.\end{tabular} ..                                                                                                             & \cellcolor[HTML]{FFFFBF}\begin{tabular}[c]{@{}l@{}}Medium-\\ High\end{tabular} \\
                                                                                     & Transp.                                            & \begin{tabular}[c]{@{}l@{}}Information related to the organization can be accessed by\\ all community members.\end{tabular}                                              & \cellcolor[HTML]{ABDDA4}High                                                   \\
                                                                                     & Privacy                                                   & \begin{tabular}[c]{@{}l@{}}Community members can independently/ unmanipulated\\ explore and experiment with solutions.\end{tabular}                                      & \cellcolor[HTML]{ABDDA4}High                                                   \\
\parbox[t]{2mm}{\multirow{-7}{*}{\rotatebox[origin=c]{90}{\multirow{2}{*}{\textbf{\begin{tabular}[c]{@{}l@{}}Collective\\ Intelligence\end{tabular}}}}}} & \begin{tabular}[c]{@{}l@{}}Free\\ express.\end{tabular} & \begin{tabular}[c]{@{}l@{}}Community members can freely express their thoughts and\\ opinions.\end{tabular}                                                              & \cellcolor[HTML]{ABDDA4}High                                                   \\ \hline
                                                                                     & Delib.                                              & \begin{tabular}[c]{@{}l@{}}Processes are in place that effectively and thoughtfully distill\\ options for decision-making from the collective intelligence.\end{tabular} & \cellcolor[HTML]{FDAE61}\begin{tabular}[c]{@{}l@{}}Low-\\ Medium\end{tabular}  \\
\multirow{-3.5}{*}{\rotatebox[origin=c]{90}{\multirow{2}{*}{\textbf{\begin{tabular}[c]{@{}l@{}}Digital\\ Democ.\end{tabular}}}}}       & Voting                                                    & \begin{tabular}[c]{@{}l@{}}Coosing (desired) system states, goals and approaches to attain\\ them fairly from the options of the deliberation process.\end{tabular}      & \cellcolor[HTML]{ABDDA4}High                                                   \\ \hline
                                                                                     & Autonomy                                                  & \begin{tabular}[c]{@{}l@{}}Community members can act as they think is best for\\ the organization.\end{tabular}                                                          & \cellcolor[HTML]{FDAE61}\begin{tabular}[c]{@{}l@{}}Low-\\ Medium\end{tabular}  \\
\parbox[t]{4mm}{\multirow{-2.8}{*}{\rotatebox[origin=c]{90}{\multirow{2}{*}{\textbf{\begin{tabular}[c]{@{}l@{}}Adap-\\ tation\end{tabular}}}}}}       & Feedback                                                  & (unsolicited real-time) feedback mechanisms are in place.                                                                                                                & \cellcolor[HTML]{ABDDA4}High  \\ \hline                                                
\end{tabular}
\label{tab:dao_framework}
\end{table}

\textit{Collective Intelligence:} A DAO is required to be open in order to leverage on a diverse set of ideas and solution. Moreover, it has to facilitate free and transparent access to the knowledge and information concerning the DAO. This includes for instance the knowledge on the functioning of deliberation and decision-making mechanisms. 
Members of the community then require privacy to reason about this information and discover/ experiment with local solutions. Finally the free expression of the newly gained insights is required such that it can be fed back into the transparent information source. 

\textit{Digital Democracy:} These diverse ideas then need to be put into a structured deliberation process that distills the best solutions (e.g. the Improvement Proposal mechanism employed by the majority of DAOs). 
In case the deliberation does not conclude on a singular solution, the voteable objects should be put into a fair voting mechanism that in turn consists of various mechanisms for vote inputs (Single-choice, Multiple-choice, etc.) and aggregation (e.g. quadratic voting).

\textit{Adaptation:} Finally, members of the organization need to be empowered to act autonomously on the implementation of those decisions which is responsive to its environment and receive feedback on their actions to adjust their actions. DAOs facilitated this by having no central authority (e.g. no legal entity such as in Bitcoin \cite{ballandies2025bitcoin}) that could censor members' actions, and utilizing token-incentives to provide feedback on beneficial actions for the DAO \cite{ballandies2021finance}.

\section{Applying the framework}
We now apply the DAO Viability Framework from Section~\ref{sec:dao_framework} to analyze the MetaDAO\footnote{The MetaDAO: https://metadao.fi (last accessed: 2025-04-12)} (Section \ref{sec:case_study}) and then illustrate how the framework can be utilized to mitigate the DAO viability challenges (Section \ref{sec:mitigating_dao_challenges}).
\subsection{Case Study: The MetaDAO}
\label{sec:case_study}
Table~\ref{tab:dao_framework} illustrates the MetaDAOs’ fulfillment of the framework’s core principles, extending the initial assessment previously presented\footnote{Futarchy - using humanity’s collective intelligence in decision-makings: https://medium.com/wihi-weather/futarchy-using-humanitys-collective-intelligence-in-decision-making-2b7dfb331082 (last accessed: 2025-04-08)}:

\paragraph{Collective Intelligence:}
The MetaDAO promotes diversity through an open Discord server with permeable entry barriers, allowing newcomers to join and immediately observe or participate in core activities. Transparency is also well-supported, as all discussion channels are public, and members can follow decision-making processes in real time. The MetaDAO further supports privacy and freedom of expression by allowing users to remain anonymous, granting them time and space to reason about issues before sharing their perspectives. 

Nevertheless, the MetaDAO could benefit from clearer onboarding materials (e.g., more prominently integrated blog articles) to help participants understand how best to contribute.

\paragraph{Digital Democracy:}
Deliberation in the MetaDAO follows a Bitcoin-inspired Improvement Proposal mechanism, supported by Discord discussions. Due to low proposal throughput and resulting inability to act efficiently, the DAO shifted away from full decentralization. \footnote{Proposal 14 - Benevolent Dictators: https://hackmd.io/@metaproph3t/SJfHhnkJC?utm\_source=preview-mode\&utm\_medium=rec, (last accessed: 2025-04-08)}. Specifically, the limited throughput and difficulty in reaching decisions led to the temporary appointment of two benevolent dictators empowered to act on behalf of the DAO.

The MetaDAO excels at the Voting principle through an innovative, market-based mechanism that continuously prices the token based on available information, incentivizing proposals that maximize token value. While it remains debatable whether such market-based voting is ideal for every possible governance context, it appears well-suited to the DAO’s primary objective of maximizing token value.

It remains unclear whether the DAO aims to achieve thorough social consensus prior to voting (via deliberation), as highlighted by the community,\footnote{For example, Durdan on Discord (21.02.2024) writes: “I think it can be partially solved via culture. If we collectively ‘shame’ creating markets [votes] without prior discussion [...], people will mostly stop doing it. Trying to do that with [...]  lots of discussion and asking for concrete feedback via polls before putting anything on chain.”} or instead relies on market mechanisms to filter out weaker proposals (via market-based voting). If social consensus prior to voting is intended, more specialized tools than the Improvement Proposal mechanism (e.g., viewpoints.xyz or pol.is) could facilitate a structured and more efficient discourse \cite{yang2025bridging}.

\paragraph{Adaptation:}
The autonomy of individuals to execute tasks for the DAO is constrained by the reliance on accepted proposals, which slows responses to urgent issues and limits creativity if every action requires voting. The implemented improvement in the form of benelovent dictators (see previous paragraph on Digital Democracy) might briefly increase action throughput compared to relying on accepted governance proposals, its scalability remains constrained by these individuals’ capacities.

The DAO provides strong feedback mechanisms: the token price reflects shifts in sentiment or organizational performance and is the deciding measure in its voting mechanism, while community reactions on Discord serve as immediate, decentralized feedback channels. 

\paragraph{Summary: }
In summary, the MetaDAO largely realizes the principles of the proposed DAO Viability Framework. It successfully cultivates collective intelligence through openness and transparency, demonstrates a robust voting mechanism for digital democracy, and integrates multiple feedback loops that foster adaptation. Improvements could be made by enhancing deliberation processes and clarifying how members can autonomously contribute, thereby further strengthening its viability as a self-organizing, complex system.

\subsection{Mitigating DAO challenges}
\label{sec:mitigating_dao_challenges}

DAO challenges emerge when the design fails to clearly define and separate the three self-organization mechanisms of the framework. A breakdown in adaptation, for instance, can lead to inactivity—an increasingly common issue in DAOs (Section~\ref{sec:dao_challenges}), as illustrated by the Meta DAO case (Section~\ref{sec:case_study}). The DAO viability framework recommends mechanisms that enhance each community member’s autonomy to act on behalf of the DAO, such as idea markets\footnote{Idea market: \url{https://github.com/wihi-labs/WHIP/blob/main/0004-idea-market.md}, (last accessed: 2025-04-03)}. This mechanism empowers any member to propose, fund, and execute tasks.

Failures in digital democracy can lead to poor decisions, while failures in collective intelligence may distort the system’s perceived state and encourage biased strategies. By conceptually distinguishing these mechanisms, system designers can explore novel solutions. For example, while most DAOs use improvement proposals for deliberation—drawing inspiration from Bitcoin—more advanced mechanisms exist \cite{yang2025bridging} and could help manage the challenges with regards to the complexities in consensus building (Section~\ref{sec:dao_challenges}).

DAO voting is often limited to single-choice binary options. Yet alternative mechanisms, like ranked-choice voting, are seen as more legitimate \cite{yang2024designing} and may improve decision-making. Additionally, the diversity in collective intelligence (Table~\ref{tab:dao_framework}) can be fostered through tailored incentives, such as minority rewards \cite{mann2017optimal}, potentially addressing the lack of experimentation in many DAOs (Section~\ref{sec:dao_challenges}).

\section{Conclusion}
\label{sec:conclusion}
DAOs are inherently complex systems that require bottom-up self-organization mechanisms for effective management. Our proposed DAO viability framework, grounded in complexity science and digital democracy, provides a structured approach to address governance and adaptation challenges of DAOs by facilitating the analysis and instantiation of three self-organization mechanisms: collective intelligence, digital democracy, and adaptation.

Current research primarily focuses on DAO voting, which aligns with the digital democracy principles of our framework (Table \ref{tab:dao_framework}). Building on this framework, future studies could delve deeper into collective intelligence and adaptation within DAOs to derive insights for enhancing DAO design. Also, linking DAO research with digital democracy and complex systems literature has proven valuable for understanding DAO functioning and could be further expanded. Finally, developing quantitative measures to assess how well DAOs fulfill self-organization principles would allow for a more objective evaluation of their viability.

\end{document}